\documentclass[prl,twocolumn,showpacs,superscriptaddress]{revtex4}
\usepackage{dcolumn}
\usepackage{bm}
\usepackage{graphicx}
\usepackage{amsmath}
\usepackage{amssymb}
\usepackage{bm}
\begin{document}
\pacs{74.70.Tx, 74.25.Fy, 72.15.Gd}
\title{The precursor state to superconductivity in CeIrIn${_5}$:
Unusual scaling of magnetotransport}
\author{Sunil Nair}
\affiliation{Max Planck Institute for Chemical Physics of Solids,
Noethnitzer Str. 40, 01187 Dresden, Germany}
\author{M.~Nicklas}
\affiliation{Max Planck Institute for Chemical Physics of Solids,
Noethnitzer Str. 40, 01187 Dresden, Germany}
\author{F.~Steglich}
\affiliation{Max Planck Institute for Chemical Physics of Solids,
Noethnitzer Str. 40, 01187 Dresden, Germany}
\author{J.~L.~Sarrao}
\affiliation{Los Alamos National Laboratory, Los Alamos, New
Mexico 87545, USA}
\author{J.~D.~Thompson}
\affiliation{Los Alamos National Laboratory, Los Alamos, New
Mexico 87545, USA}
\author{A.~J.~Schofield}
\affiliation{School of Physics and Astronomy, University of
Birmingham, Birmingham B15 2TT, United Kingdom}
\author{S.~Wirth}
\affiliation{Max Planck Institute for Chemical Physics of Solids,
Noethnitzer Str. 40, 01187 Dresden, Germany}
\date{\today}
\begin{abstract}
We present an analysis of the normal-state Hall effect and
magnetoresistance in the heavy fermion superconductor
CeIrIn${_5}$. It is demonstrated that the modified Kohler's
scaling---which relates the magnetoresistance to the Hall
angle---breaks down prior to the onset of superconductivity due to
the presence of a precursor state to superconductivity in this
system. A model-independent, single-parameter scaling of the Hall
angle governed solely by this precursor state is observed. Neither
the Hall coefficient nor the resistivity exhibit this scaling
implying that this precursor state preferentially influences the
Hall channel.
\end{abstract}
\maketitle

The variety of low temperature electronic ground states observed
in heavy fermion systems primarily arises from two competing
fundamental physical processes: the Ruderman-Kittel-Kasuya-Yosida
interaction favoring magnetic order and the Kondo effect that
screens the local moments \cite{doniach}. Of particular interest
are systems in which the magnetic order can be driven to zero
temperature. If this takes place in a continuous fashion it is
referred to as a Quantum Critical Point (QCP). The often observed
existence of unconventional superconductivity in the vicinity of
such a QCP has added to the interest in these exotic phase
transitions, as it suggests that Cooper pair formation could be
governed by the presence of (antiferro-) magnetic fluctuations
\cite{mathur}. The Ce-115 systems (of the form Ce$M$In${_5}$, with
$M$ = Co, Ir, or Rh) have proven to be an interesting playground
where manifestations of these intrinsic energy scales are
unambigously observed \cite{review}. For instance, in the ambient
pressure superconductor CeCoIn${_5}$, the QCP can be approached
with applied magnetic fields of the order of the superconducting
upper critical field $H_{c2}(0)$ \cite{paglione}. The
antiferromagnetic order observed in CeRhIn${_5}$ can be suppressed
by applying pressure of the order of 1.6 GPa which, again, results
in a superconducting ground state \cite{shishido}. In
CeIrIn${_5}$, though the superconducting regime is reasonably
separated from the (possibly metamagnetic) QCP \cite{capan},
signatures of the presence of antiferromagnetic fluctuations in
the vicinity of superconductivity have been observed \cite{zheng}.
Besides unconventional superconductivity, experimental signatures
such as the presence of line nodes in the superconducting gap
structure \cite{izawa} and anomalous magnetotransport have also
brought into focus the remarkable similarities which these systems
share with the high temperature superconducting cuprates
\cite{nakajima}.

One of the outstanding puzzles presented by these complex
materials is the changing low energy excitations of the normal
state just prior to the formation of the superconducting state. In
the cuprates for instance, it is now understood that
superconductivity is preceded by the opening of a \emph{pseudogap}
in the electronic density of states \cite{pseudogap}. Typically,
this state is associated with experimental signatures like a
deviation from the linear temperature dependence of resistivity
\cite{bucher} or a decrease in the spin-lattice relaxation rate
($1/T{_1}$) in nuclear magnetic resonance measurements
\cite{warren}, and is now considered an intrinsic energy scale of
these systems. Recently, experiments have indicated that a
precursor state to superconductivity may also exist in the Ce
based heavy fermion metals. The Ce-115 systems have been exemplary
in this aspect, with the presence of such a precursor state being
inferred from measurements like resistivity \cite{sid}, nuclear
quadrapole resonance \cite{kawasaki}, the Hall angle \cite{sunil}
and the Nernst effect \cite{bel}. In this Letter, we report on the
analysis of the normal-state magnetotransport data in
CeIrIn${_5}$. To this end, prior \cite{sunil} and new simultaneous
measurements of isothermal Hall effect and magnetoresistance in
magnetic fields $\mu_0 H \le$ 15 T and in the temperature range
0.05 K $< \textit{T} <$ 2.5 K on high-quality single crystals are
evaluated. We demonstrate that the modified Kohler's
scaling---relating the magnetoresistance to the Hall
angle---breaks down prior to the onset of superconductivity due to
a change in the Hall scattering rate. Moreover, the critical field
$H^*(T)$ of the precursor state to superconductivity alone has
been used to scale the temperature- and field-dependent Hall angle
$\theta{_H}=\cot^{-1}(\rho{_{xx}}/\rho{_{xy}})$. The fact that a
similar scaling procedure fails for the individual properties,
i.e.,\ the resistivity $\rho_{xx}$ and the Hall coefficient $R_H
=\rho_{xy}/\mu_0 H$, suggests that this precursor state
preferentially affects the Hall channel.

Fig.~1(a) depicts the isothermal Hall coefficient $|R_H|$ as a
function of applied field $H$ for different temperatures. The
sharp drop in $|R_H|$ corresponds to the onset of
superconductivity. The magnetic field dependence of the transverse
resistivity $\rho{_{xx}}$ is plotted in Fig.~1(b). Besides the
onset of superconductivity, a crossover in the sign of the
magnetoresistance corresponding to the onset of a coherent Kondo
state is visible in the high-temperature data sets. Though this
crossover from an incoherent to a coherent Kondo scattering regime
is a characteristic feature of the heavy fermion metals, many
\begin{figure}
\includegraphics[width=7cm,clip]{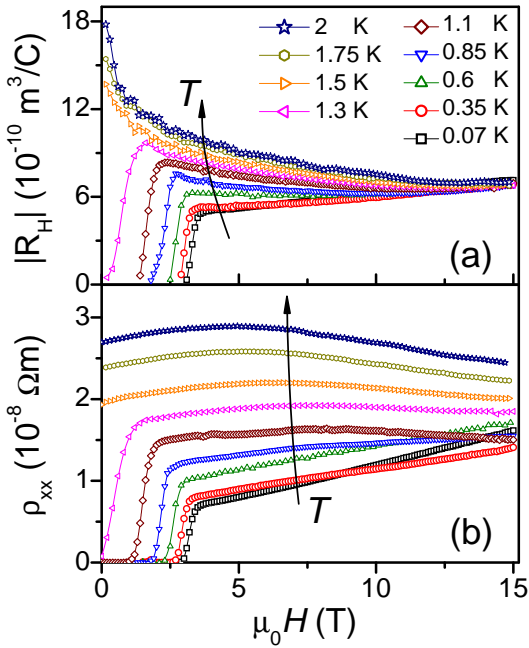}
\caption{Magnetic field dependence of (a) the Hall coefficient
$|R{_H}|$ and (b) the resistivity $\rho{_{xx}}$ measured at
selected temperatures. The sharp drop corresponds to the onset of
superconductivity.} \label{Fig1}
\end{figure}
similarities in the normal state magnetotransport of the
superconducting cuprates and the Ce-115 systems have recently come
to light. For instance, the resistivity $\rho{_{xx}}$ has a linear
temperature dependence, the Hall coefficient $R{_H}$ varies
approximately as $1/T$, and the Hall angle follows a $\cot
\theta_H \propto T{^2}$ dependence in these conceptually different
classes of materials. In the cuprates, theoretical support for
these experimental observations have relied on the rather
extraordinary idea that, in contrast to conventional metals, the
transverse Hall scattering rate ($\tau_{H}^{-1}$) in the cuprate
metals is a distinct entity as compared to the transport
scattering rate ($\tau_{tr}^{-1}$) \cite{anderson,chien}. Since
the resistivity is governed by $\tau_{tr}^{-1}$ and $R_{H}$ by the
ratio $\tau_{H}$/$\tau_{tr}$, it follows that $\cot\theta{_H}$ is
a manifestation of the transverse relaxation rate $\tau_{H}^{-1}$
alone. In conventional metals (with an isotropic single scattering
rate), the magnetoresistance ($\rho_{xx}(H) -
\rho_{xx}(0))$/$\rho_{xx}(0)$ arising due to the orbital motion of
charge carriers is known to scale as a function of
$H/\rho_{xx}$(0) (Kohler's rule \cite{pippard}). A natural
consequence of the presence of two scattering rates was the
re-formulation of this scaling rule to relate the transverse
magnetoresistance with the Hall angle \cite{harris}. This scaling
of the form ($\Delta\rho{_{xx}}/\rho_{xx}(0)) \propto
\tan^2\theta_H$ has been successfully applied to magnetotransport
data in both the cuprates as well as in all the Ce-115 compounds.
In CeIrIn${_5}$ for instance, it was recently demonstrated that
this scaling works in a wide temperature range down to about 2 K
\cite{nakajima1}, but this study did neither extend down to the
precursor nor the superconducting regime [cf.\ Fig.\
\ref{Fig2}(b)]. The vital question remained whether the
superconducting condensate emerges from within the phase space
\begin{figure}
\includegraphics[width=8.5cm,clip]{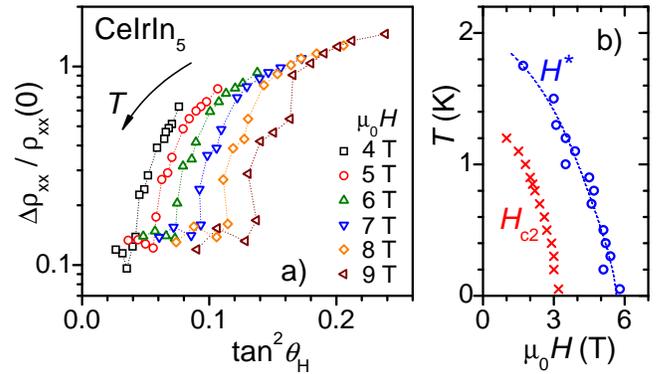}
\caption{(a) Magnetoresistance vs.\ squared tangent of the Hall
angle $\theta{_H}$ revealing strong deviations from the modified
Kohler's rule. Here, temperature is an implicit parameter as
indicated. (b) Part of the $H-T$ phase diagram of CeIrIn${_5}$
exhibiting the boundary of superconductivity and its precursor
state (marked by $H{_{c2}}$ and $H^*$, respectively).}
\label{Fig2}
\end{figure}
where this scaling is obeyed. Fig.\ \ref{Fig2}(a) exhibits the
modified Kohler's scaling as determined from our new
magnetotransport data. Clearly, the scaling procedure mentioned
above is \emph{not} applicable down to the lowest accessible
temperatures. This observation is in line with the inference that
the formation of the superconducting condensate in CeIrIn${_5}$ is
preceded by a precursor state as determined by a change in the
Hall mobility.

\begin{figure}
\includegraphics[width=7cm,clip]{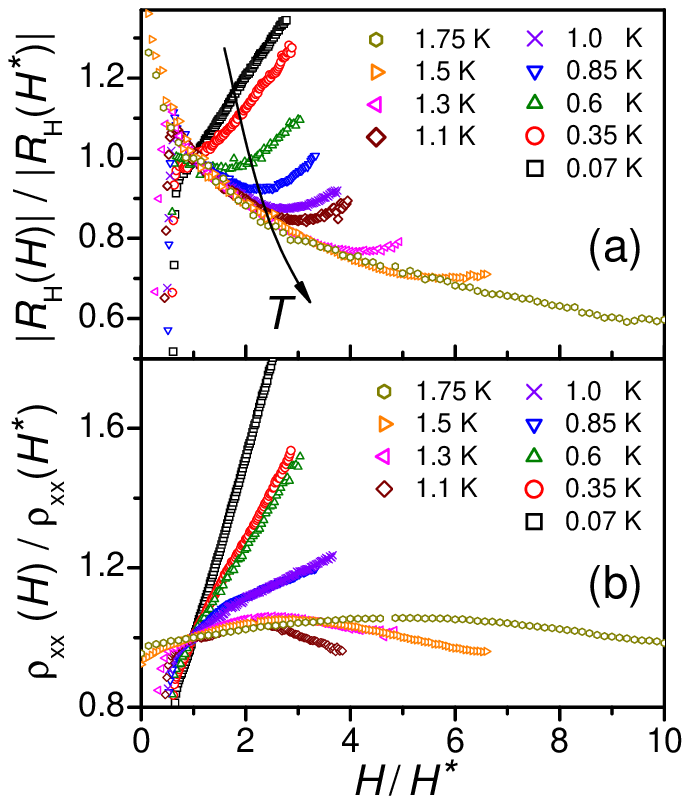}
\caption{(a) The scaled field dependences of the  the normalized
Hall coefficient $|R{_{H}}|(H)/|R{_{H}}|(H^*)$ and (b) the
normalized resistance $\rho{_{xx}(H)}/\rho{_{xx}}(H^*)$ plotted as
a function of the normalised field ($H/H{^*}$) for CeIrIn${_5}$.}
\label{Fig3}
\end{figure}

In the heavy fermion metals, the crystal electric field and the
single ion Kondo effect provide two fundamental energy scales that
crucially influence its physical properties. An additional energy
scale of importance \cite{aeppli,rossat} is related to the
intersite coupling between the local moments due to the
Ruderman-Kittel-Kasuya-Yosida interaction. In this context, it is
important to clarify whether this precursor state to
superconductivity in CeIrIn${_5}$ represents an intrinsic energy
scale of the system, and to discern the manner in which it
influences the normal-state magnetotransport. One powerful tool of
identifying intrinsic energy scales in strongly correlated systems
is the quest for universal trends of, and relationships between,
measured physical quantities. In the heavy fermion systems, early
attempts to scale physical properties using a single energy
scaling parameter met with only limited success \cite{scaling}.
However, in the cuprates it has been demonstrated
\cite{hwang,wuyts,takemura} that a single-parameter scaling of
experimental data was possible by using the energy scale of the
pseudogap alone. By normalizing any measured electrical or thermal
transport quantity $f(x)$ along with its variable $x$ by the
corresponding values at the onset of the pseudogap [$f(x^*)$ and
($x^*$), respectively] the measured data could be made to collapse
into a single universal curve. Thus, the scaling is of the form
$f(x)/f(x^*) \propto F(x/x^*)$. Consequently, the normalized Hall
coefficient ($|R_{H}(H)|/|R_{H}(H^*)|$) is plotted as a function
of normalized field ($H / H^*$) for CeIrIn${_5}$ in Fig.\
\ref{Fig3}(a). Here, the values $H^*$ [Fig.\ \ref{Fig2}(b)] were
determined earlier from the change in the Hall mobility
\cite{sunil}. Fig.\ \ref{Fig3}(b) shows the equivalent scaling for
the magnetoresistance, i.e.\ $\rho_{xx}(H)/\rho{_{xx}}(H{^{*}})$
as a function of $H/H^*$. Interestingly, neither $R_{H}(H)$ nor
$\rho{_{xx}}(H)$ scale onto a universal curve implying that both
of these quantities have significant contributions which are not
scaling-invariant.

\begin{figure}
\includegraphics[width=7cm,clip]{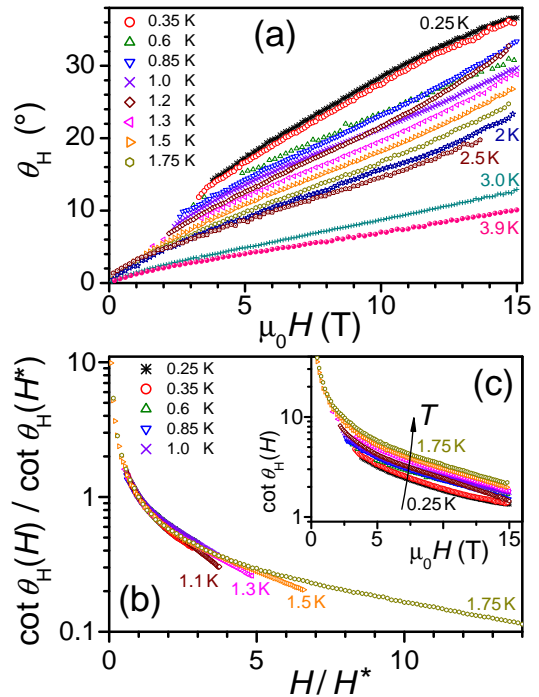}
\caption{(a) Field dependence of the Hall angle $\theta{_H}$. (b)
Scaling of both $\cot\theta{_H}$ and field $H$ with respect to
$H^*$ reveals a collapse of the data into a single generic curve.
The inset (c) shows the field dependence of some unscaled data
sets.} \label{Fig4}
\end{figure}

A priori, there is no simple explanation of the nature of these
{\em non}-scaling-invariant contributions to $R_{H}$ and
$\rho{_{xx}}$. One possibility which cannot be ruled out is the
influence of disorder in these systems, with impurity scattering
not being scaling-invariant. This problem can be circumvented by
the analysis of the Hall angle. From prior work on the cuprates it
is known that the cotangent of the Hall angle (which is directly
related to the charge carrier mobility) is a quantity of basic
interest \cite{chien}. It has been shown that $\cot\theta_H$
follows a $T{^2}$ dependence, independent of the extent of
impurity substitution as well as the charge carrier density
\cite{yakabe}. This relative insensivity of $\cot\theta{_H}$ to
material properties (which is related to the fact that it does
{\em not} depend on $\tau_{tr}$) has led to the conjecture that it
is an even more fundamental property than $R_{H}$. Moreover,
deviations from $\cot\theta{_H} \propto T^2$ have been used to
identify the onset of the pseudogap state in the cuprates
\cite{abe}. Fig.\ \ref{Fig4}(a) presents the isotherms of the
field dependent Hall angle as measured in CeIrIn${_5}$, indicating
that $\theta{_H}$ is quasi-linear as a function of $H$. If the
different electronic states in CeIrIn${_5}$ are manifestations of
a change in the geometry of the Fermi surface, this should be
visible in the field dependence of the Hall angle which measures
the effective deflection of charge carriers in the material by the
applied magnetic field. However, the lack of any observable
features at fixed values of $\theta{_H}$ suggests that there is no
abrupt change in the geometry of the Fermi surface, at least in
the range of our measurements. Here, it is emphasized that
$\theta{_H}$ attains a value of more than 30$^\circ$ at large
fields which is substantially larger than what is commonly
observed in the cuprates. In line with the earlier analysis, the
normalized Hall angle $\cot\theta{_H}(H)/ \cot\theta{_H}(H^*)$ is
plotted as a function of normalized field $H/H^*$ in Fig.\
\ref{Fig4}(b). For comparison, the unscaled data are shown in the
inset Fig.\ \ref{Fig4}(c). Only in the former case a good scaling
behavior is obtained, an observation which is remarkable in view
of the fact that such a scaling procedure was found to be
ineffective for both $\rho_{xx}(H)$ and $R{_H}(H)$, Fig.\
\ref{Fig3}. This scaling of $\cot\theta{_H}$ unambiguously implies
that the precursor state observed in CeIrIn${_5}$ represents an
intrinsic energy scale of the system which influences the
magnetotransport in a substantial region of the field-temperature
phase space. Note that the scaling of the critical field $H^*(T)$
of the precursor state with the superconducting critical field
$H_{c2}(T)$ suggests that they may arise from the same underlying
mechanism \cite{sunil}. This provides a natural link between the
normal-state properties of CeIrIn${_5}$ and the superconductivity
in this system.

The fact that scaling is observed in $\cot\theta_{H}$ clearly
suggests that the precursor state is primarily associated with the
transverse Hall scattering rate $\tau_{H}^{-1}$. However, it would
be erronous to conclude that the precursor state is associated
\emph{only} with $\tau_{H}^{-1}$, since this state is also
identified by a subtle feature in the magnetoresistance
\cite{sunil}. Nevertheless, the lack of scaling in both
$\rho{_{xx}}$ and $R_{H}$ suggests that the magnetic field
seemingly influences $\tau_{H}^{-1}$ preferentially as compared to
$\tau_{tr}^{-1}$. Interestingly, this is also in agreement with
prior results on underdoped cuprates where it was suggested that
the formation of a pseudogap primarily affects the Hall channel,
and has little effect on the diagonal conductivity \cite{Xu}.
Moreover, the observation of scaling \emph{only} in
$\cot\theta{_H}$ re-emphasizes the presence of two distinct
scattering processes which selectively influence the resistivity
and the Hall angle in this heavy fermion metal. The observed
anisotropy in the magnetic field response of the scattering rates
may arise as a consequence of coupling of the quasiparticles to
incipient antiferromagnetic fluctuations \cite{pines,kon}. Such a
coupling might then renormalize the scattering rates along
different directions of the Fermi surface. There exists a body of
work to imply that this might indeed be the case in the Ce-115
systems. For instance, investigations of the angular-dependent
resistivity in CeCoIn${_5}$ have indicated the presence of two
distinct regimes in their magnetic field dependences, separated by
a critical angle $\theta{_{c}}$, which in turn is governed by the
intrinsic anisotropy \cite{hu}. Moreover, recent thermal
conductivity measurements indicated that the superconducting gap
of CeIrIn${_5}$ may have a $d{_{x{^2}-y{^2}}}$ symmetry: a
signature that the superconductivity is strongly influenced by the
presence of antiferromagnetic fluctuations \cite{kasahara}. These
fluctuations are themselves inferred to be anisotropic in nature,
with the magnetic correlation length along the basal plane being
larger than along the $c$ axis, $\xi_{ab} > \xi_{c}$ \cite{zheng}.
The two corresponding scattering rates appear to be influenced by
the low-lying precursor state in a disparate fashion. In
CeIrIn${_5}$, it has been observed that $\cot\theta_H$ increases
anomalously in the precursor state \cite{sunil}. Since
$\cot\theta{_H} = 1/\omega{_c}\tau_H$ this suggests that
$\tau_H^{-1}$ is \emph{enhanced} in the precursor
state---provided, of course, that the effective mass $m^*$ remains
constant. This is in contrast to observations in the cuprates as
well as in the related system CeCoIn${_5}$ \cite{sid}, where it
was found that $\tau_{tr}^{-1}$ \emph{reduces} at the onset of the
precursor state.

In summary, the analysis of the normal-state magnetotransport in
CeIrIn${_5}$ reveals that the modified Kohler's plot (relating the
magnetoresistance to the Hall angle) breaks down prior to the
onset of superconductivity, presumably due to the presence of a
precursor state to superconducivity in this regime of the $H$--$T$
phase space. Moreover, the Hall angle obeys a single-parameter
scaling unambiguously governed by this precursor state. The
absence of scaling in $R{_H}$ and $\rho{_{xx}}$ is clearly
indicative of the presence of two distinct scattering times, in
similarity with observations in the cuprate superconductors. This
could very possibly be a generic feature of many heavy fermion
superconductors. The fact that \emph{only} the Hall angle is
scaled by the precursor state also implies that this state
preferentially influences the Hall channel and has a relatively
weaker influence on the resistivity. It is imperative to map the
evolution and symmetry of both the superconducting as well as the
precursor state by more direct probes, e.g.,\ in order to
formulate a theoretical basis for the observed phenomena.

The authors thank A. Gladun for useful discussions. S.N.\ is
supported by the Alexander von Humboldt foundation. S.W.\ is
partially supported by the EC through CoMePhS 517039. Work at Los
Alamos was performed under the auspices of the U.S.\ Department of
Energy/Office of Science. Work at Dresden was supported by DFG
research unit 960. AJS acknowledges support of the MPI PKS,
Dresden where part of his work was done.

\end{document}